\documentclass[aps,prb,twocolumn,showpacs]{revtex4}
\usepackage{amsmath}
\usepackage{graphicx}

\newcommand{\al}{\alpha}
\newcommand{\be}{\beta}
\newcommand{\g}{\gamma}
\newcommand{\de}{\delta}
\newcommand{\e}{\epsilon}

\newcommand{\thi}{\theta}

\newcommand{\la}{\lambda}

\newcommand{\s}{\sigma}

\newcommand{\w}{\omega}
\newcommand{\G}{\Gamma}
\renewcommand{\S}{\Sigma}

\newcommand{\ex}[1]{e^{#1}}

\newcommand{\round}[1]{\left({#1}\right)}
\newcommand{\square}[1]{\left[{#1}\right]}
\newcommand{\cvec}[2]{\round{\begin{array}{c}#1\\#2\end{array}}}
\newcommand{\mat}[4]{\round{\begin{array}{cc}#1&#2\\#3&#4\end{array}}}

\begin{document}

\title{Evolution of the single-hole spectral function across a quantum
phase transition in the anisotropic-triangular-lattice antiferromagnet}

\author{So Takei$^{1}$}
\author{Chung-Hou Chung$^{1}$}
\author{Yong Baek Kim$^{1,2}$}
\affiliation{$^{1}$Department of Physics, University of Toronto, Toronto,
Ontario M5S 1A7, Canada\\
$^{2}$School of Physics, Korea Institute for Advanced Study, Seoul 130-722,
Korea}
\date{\today}

\begin{abstract}
We study the evolution of the single-hole spectral function when the ground
state of the anisotropic-triangular-lattice antiferromagnet changes from
the incommensurate magnetically-ordered phase to the spin-liquid state.
In order to describe both
of the ground states on equal footing, we use the large-$N$ approach where the
transition between these two phases can be obtained by controlling the
quantum fluctuations via an 'effective' spin magnitude. Adding a hole into these
ground states is described by a $t$-$J$ type model in the slave-fermion
representation. Implications of our results to possible future ARPES experiments
on insulating frustrated magnets, especially Cs$_2$CuCl$_4$, are discussed.
\end{abstract}


\maketitle

\section{Introduction}

The physics of geometrically frustrated quantum magnetic systems has been an
arena for discovery of new and interesting phenomena both on the theoretical and
experimental fronts. Much of the effort has been sparked by Anderson's proposal
of the exotic resonating valence bond (RVB) spin-liquid phase
\cite{anderson73,anderson87} and the belief that frustrated systems are
likely to possess RVB spin-liquid-like
ground states in two dimensions. Theoretical efforts in the past have explored
a variety of two-dimensional frustrated systems\cite{misguich03,sachdev03,tri}.
The discovery of superconductivity\cite{takada03} in
Na$_{0.35}$CoO$_2\cdot$1.3H$_2$O has
also raised the hope that the RVB could be the cause of superconductivity in
certain frustrated magnetic systems.

Recent theoretical works on the insulating anisotropic-triangular-lattice antiferromagnet
\cite{moessner01,chung01,chung03,zhou02,essler01} have been
particularly motivated by the experimental discovery of the fractionalized spin-liquid
phase in Cs$_2$CuCl$_4$\cite{coldea03,coldea01}, where spin-1/2 moments of Cu
antiferromagnetically interact with each other in the layers of anisotropic
triangular lattices. The experiment has shown that, with a sufficiently high
magnetic field within its ordering plane, the compound looses its long-ranged
cycloid order and enters the spin-liquid phase\cite{coldea03,coldea01}.
In this phase, the dynamical spin
structure factor shows a broad profile characteristic of fractionalization of
spin-1 magnons into a pair of spin-1/2 spinons\cite{coldea03,coldea01}.

A useful description of this transition can be obtained in the large-$N$
approach\cite{chung01,read90,sachdev91,read91,sachdev91-2,sachdev92,marston01,sondhi01,
bernier03}
to the antiferromagnetic Heisenberg model on the anisotropic triangular lattice.
Here the spin SU(2) symmetry is generalized to Sp($N$) and the large-$N$ limit
of the model is investigated. In this approach, the quantum fluctuations are
controlled by the parameter $\kappa$ that is the Sp($N$) analog of the spin
magnitude. The resulting phase diagram contains both the incommensurate
long-ranged order (LRO) phase\cite{joli90}
(this corresponds to the cycloid order) and
the spin-liquid state\cite{z2} in the large and small
$\kappa$ limits, respectively\cite{chung01,sachdev92}.
Thus one may expect that $1/\kappa$ plays a  
role similar to the in-plane magnetic field in Cs$_2$CuCl$_4$.
Indeed the spin structure factor of the spin-liquid state computed in the
large-$N$ theory\cite{chung03,zhou02} looks very similar to that observed in the
neutron scattering experiments on Cs$_2$CuCl$_4$\cite{coldea03,coldea01}.

To this date, experimental evidence of the RVB spin-liquid phase in the
anisotropic triangular lattice has been obtained only in neutron scattering
experiments on Cs$_2$CuCl$_4$\cite{coldea03,coldea01}. Thus it
is desirable to have independent experimental confirmations of these results.
One such experiment may be the angle-resolved photoemission spectroscopy (ARPES)
that measures the electron spectral function. In particular, the result of ARPES on
an insulating compound at half-filling corresponds to measuring the
single-hole spectral function in a given ground state of the insulator\cite{wells95}.

In this paper, we investigate the evolution of the single-hole spectral function
across the transition between the incommensurate LRO phase and the spin-liquid
phase. Previously the single-hole spectral function in the LRO phase was investigated
by taking into account the interaction of the hole with the spin
waves\cite{vojta99,azzouz96,kane89,marsiglio91}.
The well-defined low-energy quasiparticle peak was found in these studies.
This is due to the fact that small low-energy density of states of the magnons
leads to only a weak perturbation to the quasiparticle coherence at low energy.
While this calculation is enough to describe the spectral function
deep inside the ordered phase, an additional ingredient may have to be
considered close to the transition toward the nearby spin-liquid state
where the spinons are deconfined\cite{z2}. The confinement of the spinons in
the LRO phase is a low energy phenomenon and it is expected that the
confinement energy scale will be very small near the transition.
Therefore, it is expected that there will be spinon excitations at
intermediate to high energies near the transition and they may affect
the coherence of the hole motion. For example, the neutron scattering
experiments on the magnetically ordered phase of
Cs$_2$CuCl$_4$ found the spinon continuum as well
as the magnon peak, implying that this
ordered phase is very close to the transition\cite{coldea03,coldea01}.
Indeed the dominant interaction near the transition may be the spinon-hole
interaction since the magnons are relatively suppressed close to the transition.

We use the results of the large-$N$ theory to describe the transition
between two ground states at half-filling. In this formulation, as will be
explained later, the spinons appear at the mean field level even in
the LRO phase\cite{chung01,read90,sachdev91,read91,sachdev91-2}.
Even though this is an artifact of the mean field theory and the low energy
spinons are confined due to the fluctuation effects beyond the mean field
theory\cite{read90,sachdev91,read91,sachdev91-2},
we can still use these spinons at intermediate to high energies to mimic the
situations near the transition as described above.
The motion of the single-hole is incorporated via a $t$-$J$ type model
in the slave fermion representation\cite{kane89}.
Our computation takes into account the multiple interactions between the motion
of the hole and the spin background, but does not include the vertex corrections.
In principle, the latter approximation could lead to an underestimation of the coherence in
the motion of the hole. Comparison to the exact diagonalization study\cite{lauchli03},
however, suggests that the latter approximation does not change the overall
structure of the spectral function in a significant fashion.

Our results indicate that the single-hole spectral function
still shows a well-defined low-energy quasiparticle peak in the
incommensurate LRO phase
with a visible dispersion in momentum space. This is due to the fact
that the density of states of spinons becomes smaller as the energy scale is
decreased and, as a result, does not strongly disturb the low energy motion
of the hole. As the transition to the spin-liquid state is approached,
the spectral weight of the quasiparticle peak is decreased and the incoherent
continuum at intermediate to high energies grows. Eventually the quasiparticle
peak disappears at the transition. In the spin-liquid state,
the single-hole spectral function only shows a broad incoherent continuum with
little dispersion, which can be interpreted as a consequence of
spin-charge separation.  

The rest of the paper is organized as follows.
In section II, we describe the formalism used to compute the single-hole
spectral function and explain the approximation schemes used in this work.
In section III, the numerical results on the single-hole spectral function
in various phases are discussed. Finally, we summarize our results and
conclude in section IV.

\section{Formulation of the problem}

At half-filling, the quantum ground states of the Mott insulator on the anisotropic 
triangular lattice are described by the Heisenberg model:
\begin{equation}
H_J = J_1 \sum_{\langle i,j \rangle} {\bf S}_i \cdot {\bf S}_j
+ J_2 \sum_{\langle \langle i,j \rangle \rangle} {\bf S}_i \cdot {\bf S}_j,
\label{heisenberg}
\end{equation}
where the first and second terms represent the nearest and the next-nearest
neighbour Heisenberg couplings, respectively (see Fig.1).
The motion of the single hole in a given ground state of the Mott
insulator can be studied by adding the kinetic energy terms of the
electrons;
\begin{equation}
H_t = -t_1 \sum_{\langle i,j \rangle} (c^{\dagger}_{i\sigma}c_{j \sigma} + h.c.)
- t_2 \sum_{\langle \langle i,j \rangle \rangle}
(c^{\dagger}_{i\sigma}c_{j\sigma} + h.c.),
\label{hopping}
\end{equation}
where $c^{\dagger}_{i\sigma}$ is the creation operator of the electrons
with spin $\sigma = \uparrow, \downarrow$ (repeated indices are summed)
and it is assumed that there is no doubly occupied site
in the Hilbert space to take into account the strong correlation inherent
in the Mott insulator.

\begin{figure}
\begin{center}
\scalebox{0.5}{\includegraphics*[198,488][437,730]{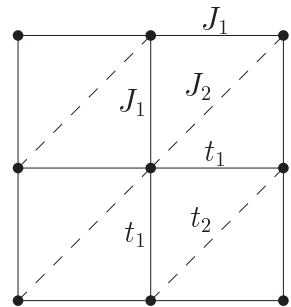}}
\caption{\label{fig:trlat}  Anisotropic triangular lattice with two types of bond.
The lattice is modelled by a square lattice with an additional interaction along one
 of the diagonals. The nearest neighbour coupling is associated with subscript 1 and
 the next-nearest neighbour coupling is associated with subscript 2.}
\end{center}
\end{figure}

\subsection{Ground States of the Mott Insulator via the
Large-$N$ Sp($N$) Heisenberg Model}

A useful formulation of the Heisenberg model in Eq.\ref{heisenberg}
can be obtained from
a bosonic representation of the SU(2) spin via
$S^a_i = {1 \over 2} b^{\dagger}_{i\sigma} \tau^a_{\sigma \sigma'}
b_{i\sigma'}$, where $\tau^a$ ($a=x,y,z$) are the Pauli matrices
and $b_{i\sigma}$ is the canonical bosonic operator.
In addition, the constraint $b^{\dagger}_{i\sigma} b_{i\sigma}=
n_b=2S$ must be imposed at each site, where $S$ is the magnitude of the spin.
In order to study this model on frustrated lattices in a controlled fashion and
to obtain results that are non-perturbative in the coupling constants,
we consider the Sp($N$) generalization\cite{read90,sachdev91,read91,sachdev91-2}
of the physical spin SU(2) $\simeq$ Sp(1).
This can be achieved by introducing $N$ flavours of bosons at each site,
$b_{i\alpha}$, where $\alpha=(n,\sigma)$ and $n=1,...,N$ is the flavour index.  
Here again the boson number at each site,
$n_b = b^{\dagger}_{i\alpha} b_{i\alpha}$, must be fixed.
The resulting Sp($N$) Heisenberg model is given by\cite{read90,sachdev91,read91,sachdev91-2}
\begin{equation}
H_{{\rm Sp}(N)}=-\sum_{i>j}\frac{J_{ij}}{2N}
(\mathcal{J}_{\al\be}b^{\dag}_{i\al}b^{\dag}_{j\be})
(\mathcal{J}_{\g\de}b_{i\g}b_{j\de}),
\label{Hspn}
\end{equation}
where $J_{ij}=J_1$ and $J_2$ on the nearest and next-nearest neighbour
links, respectively. Here ${\cal J}_{\alpha\beta} = - {\cal J}_{\beta\alpha}$ is
a generalization of the SU(2) antisymmetric tensor
$\epsilon_{\sigma\sigma'}=-\epsilon_{\sigma'\sigma}$ and is
given by a $2N \times 2N$ matrix that has $N$ copies of $\epsilon$
along its center block diagonal. The matrix is zero elsewhere.
  
We consider the large-$N$ limit of this model at a fixed boson density
per flavour, $\kappa=n_b/N$. Notice that $\kappa = 2S$ when $N=1$.
Thus $\kappa$ is the Sp($N$) analog of $2S$ and controls the strength of
quantum fluctuations. The large-$N$ mean-field theory of this
model has been analyzed\cite{chung01} and the phase diagram at large-$N$ has been
obtained as a function of $\kappa$ and $J_2/(J_1+J_2)$
(see Fig.2)\cite{chung01}. For large values of $\kappa$, quantum fluctuations
are small so that the LRO states
arise by breaking the global Sp($N$) symmetry. In the small $\kappa$
limit, strong quantum fluctuations lead to paramagnetic states with
only short-ranged order (SRO). In the phase diagram of Fig.2, the ordering
wavevector of the LRO states (and the corresponding SRO phases) are
labeled by the ordering wavevector ${\bf q}=(q_1,q_2)$ in two dimensions.
It has been known that $(\pi,\pi)$ SRO phase develops the valence-bond-solid
order upon inclusion of singular fluctuations beyond the mean-field theory
and the Berry phase effect\cite{sachdev03,chung01}.  
In this paper, we will consider the transition between the incommensurate
$(q,q)$ LRO and the $(q,q)$ SRO phases that respectively correspond to the
cycloid-ordered and the translationally-invariant spin-liquid phases
discovered in the experiment involving Cs$_2$CuCl$_4$\cite{coldea03,coldea01}.  

\begin{figure}
\begin{center}
\includegraphics[bb=0 0 495 393,clip,scale=0.45]{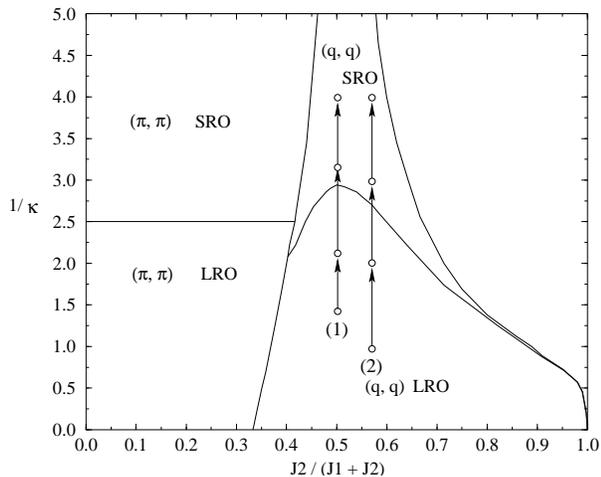}
\caption{\label{fig:triphase}  The zero temperature phase diagram\cite{chung01}
of the Sp($N$) Heisenberg antiferromagnet on the anisotropic triangular lattice
at large $N$. $1/\kappa$ represents the strength of quantum
fluctuations. Single hole spectral functions are evaluated at each of the ground states
denoted by the circles along paths (1) and (2). See text for details.}
\end{center}
\end{figure}

Now, for future reference, some explanation of the large-$N$ mean-field
theory is in order. The presence of the LRO can be described by the condensation
of bosons, $\langle b_i^{n\s}\rangle=\sqrt{N}\de_1^n x^\s_i$, where
the $c$-number field $x^{\sigma}_i$ quantifies the condensate.
Thus, in the LRO phase, it is convenient to parametrize the bosonic fields as
\begin{equation}
b_i^{n\s} = \round{\begin{array}{c}\sqrt{N}x_i^\s\\b_i^{m\s}\\\end{array}},
\label{par}
\end{equation}
where $m=2,...,N$. On the other hand, none of the bosons is condensed in
the SRO phase.
After decoupling the quartic boson interaction in Eq.\ref{Hspn}
by the Hubbard-Stratonovic fields $Q_{ij}$, the mean-field Hamiltonian
can be obtained as\cite{read90,sachdev91,read91,sachdev91-2,sachdev92}
\begin{eqnarray}
H_{MF}&=&\sum_{i>j}\left[\frac N2J_{ij}Q_{ij}\right.\cr
&-& \left. \frac{\e_{\s\s'}}2J_{ij}Q_{ij}\round{Nx^\s_ix^{\s'}_j+\sum_{m=2}^{N}
b_i^{m\s}b_j^{m\s'}}+h.c. \right ] \cr
&+&\sum_i\la_i\round{N|x_i^\s|^2+
\sum_{m=2}^{N}b_{i,m\s}^{\dag}b_i^{m\s}-n_b},
\label{HMF}
\end{eqnarray}
where $Q_{ij}={1 \over N} \langle {\cal J}_{\alpha \beta} b^{\dagger}_{i\alpha}
b^{\dagger}_{j\beta} \rangle$ at the saddle of the corresponding action.
Here the last term of Eq.\ref{HMF} takes care of the constraint on the boson
number at each site and $\lambda_i$ is the Lagrange multiplier.

At the mean-field level, $Q_{ij}=Q_1$ and $Q_2$ on the nearest and the
next-nearest neighbour links, and $\lambda$ is a site-independent constant.
In the LRO phase, the boson condensate fraction has the form
\begin{equation}
x_{i\uparrow} = x e^{i {\bf Q}\cdot {\bf r}_i}, \hskip 0.3cm
x_{i\downarrow} = -i x^* e^{-i {\bf Q}\cdot {\bf r}_i},
\end{equation}
where ${\bf q}=2{\bf Q}$ is the magnetic ordering wavevector.
In the mean-field theory, $Q_1, Q_2, x, \lambda$ and {\bf Q} have to be
determined self-consistently for a given $\kappa=n_b/N$.

The part of the Hamiltonian for the uncondensed bosons or spinons, 
$b_{im\sigma}$ ($m=2,...,N$), can be diagonalized using the Bogoliubov transformation:
\begin{equation}
H'_{MF}=\sum_{\bf k} \sum_{m=2}^N \e_{\bf k}
\left ( \al^\dag_{{\bf k}m}\al_{{\bf k}m}+
\be^\dag_{{\bf k}m}\be_{{\bf k}m} \right ),
\label{HMFfinal}
\end{equation}
where
\begin{eqnarray}
\e_{\bf k} &=& \sqrt{\la^2-|\g_{\bf k}|^2} \cr
\g_{\bf k} &=& i\square{J_1Q_1(\sin k_x+\sin k_y)+J_2Q_2\sin(k_x+k_y)} \cr
&=&\left |\g_{\bf k} \right| e^{i\thi_\mathbf{k}}.
\label{gamma}
\end{eqnarray}
The Bogoliubov quasiparticle operators, $\al_{{\bf k}m}$ and
$\be_{{\bf k}m}$, are given by
\begin{equation}
\cvec{\al_{{\bf k}m}}{\be_{-{\bf k}m}^\dag}
=\mat{u_{\bf{k}}}{v_{\bf{k}}}{v^*_{\bf{k}}}{u_{\bf{k}}}
\cvec{b_{{\bf k}m\uparrow}}{b^\dag_{-{\bf k}m\downarrow}},
\label{Bogospin}
\end{equation}
where
\begin{equation}
u_{\bf k}=\frac{1}{\sqrt{2}}\sqrt{\frac{\la}{\e_{\bf k}}+1},
\hskip 0.3cm
v_{\bf k}=\frac{\ex{-i\thi_{\bf k}}}{\sqrt{2}}
\sqrt{\frac{\la}{\e_{\bf k}}-1}.
\end{equation}
Notice that, in the SRO phase, there are no condensed bosons and
all the spinons are described by Eq.\ref{HMFfinal}.

It has been known that the presence of unconfined spinons is an artifact
of the large-$N$ mean-field theory and these spinons are, in principle,
confined at low energies due to fluctuations beyond the
mean-field theory\cite{read90,sachdev91,read91,sachdev91-2}
Thus the spinons cannot be seen deep inside the LRO phases.
The confinement scale below which the confinement occurs, however, can
be very small when the transition to the spin-liquid phase
is approached. Thus we expect that there exist 'almost'-deconfined
(or loosely-confined) spinons at intermediate to high energy scales near
the transition. At the transition point, this energy scale should be
zero because the spinons are deconfined in the spin-liquid phase.
In fact, the experiment on Cs$_2$CuCl$_4$ reveals the
existence of both of the magnon peak and the spinon continuum in the
cycloid-ordered phase, indicating that this system is very close to
the transition toward the spin-liquid phase\cite{coldea03,coldea01}.
While the confinement phenomenon cannot be captured by the mean-field
theory, the existence of the deconfined spinons at the mean-field level
can be used to phenomenologically describe the existence of 'almost'-deconfined
spinons near the transition. From now on, we will adopt this
picture and examine the effect of the 'almost'-deconfined
spinons on the motion of the holes.

\subsection{Motion of Single Hole and the Spectral Function}

In order to describe the motion of a single hole in the large-$N$ limit,
we use the slave-fermion representation of the electron
operator with the flavour index:
\begin{equation}
c^{\dag}_{i,n\sigma} = f_i \ b^{\dag}_{i,n\sigma},
\end{equation}
where $f_i$ represents the fermionic charge degree of freedom and
$b_{i,n\sigma}$ the bosonic spin degree of freedom of the electrons.
  
The single-hole spectral function can be computed from the single-hole
Green's function:
\begin{eqnarray}
G_{h,\s}({\bf r},\tau)
&=& {1 \over N} \sum_n G_{h,n\s}({\bf r},\tau) \cr
&=& -{1 \over N} \sum_n \langle
T_{\tau}c^{\dag}_{n\s}({\bf r},\tau)c_{n\s}(0,0) \rangle,
\label{hp}
\end{eqnarray}
where $T_{\tau}$ is the time-ordering operator of
the imaginary time. Notice that the physical
hole Green's-function is defined as the average of the
Green's functions in different flavour channels.
Since no flavour degree of freedom is special on physical
grounds, this is a legitimate definition of the hole
Green's function that turns out to be particularly
useful for our purposes.
 
After analytic continuation, the spectral function
can be obtained from
\begin{equation}
A_{h,\sigma} ({\bf k},\omega)=-\frac 1\pi\mbox{Im}[G_{h,\s}({\bf k},\omega+i\delta)],
\label{shsf}
\end{equation}
where $G_{h,\s}({\bf k},\omega)$ is the single-hole Green's function
in Fourier space.

At the mean-field level, the single-hole Green's function can be
written as
\begin{eqnarray}
G_{h,\s}({\bf r},\tau) \approx
-{1 \over N} \sum_n G_f({\bf r},\tau)G_{b,n\s}({\bf r},\tau),
\label{hpdecouple}
\end{eqnarray}
where
\begin{eqnarray}
G_f({\bf r},\tau) &=&
-\langle T_{\tau}f({\bf r},\tau)f^{\dag}(0,0) \rangle \cr
G_{b,n\s}({\bf r},\tau) &=&
-\langle T_{\tau}b^{\dag}_{n\s}({\bf r},\tau)b_{n\s}(0,0) \rangle.
\end{eqnarray}
In Fourier space, we get
\begin{multline}
G_{h,\s}({\bf k},\omega)\\
=-\frac{1}{N} \sum_n \sum_{{\bf q}}
\int d\nu \ G_{b,n\s}({\bf k-q},\omega -\nu)G_f({\bf q},\nu).
\label{hpf}
\end{multline}
Thus the single-hole Green's function is given by the convolution of the fermionic
and the bosonic Green's functions.  

In this work, we will assume that the injection of the single-hole does not affect
the underlying ground state at half-filling. In this case, it may be sufficient
to use the bare bosonic Green's-function, $G^0_{b,n\sigma}$, in the absence of the hole
\cite{kane89}. On the other hand, the fermionic degree of freedom of the hole will
be very much affected by the presence of the excitations in the underlying ground state.
We will compute the renormalized fermionic Green's function using the self-consistent
Dyson equation\cite{kane89,marsiglio91},
where the non-crossing diagrams are summed. Possible effect of the
vertex corrections in the fermionic Green's function and Eq.\ref{hpdecouple}
will be discussed later.

\subsubsection{Computation of the Fermion Green's Function}

The proper large-$N$ generalization of the kinetic energy terms of the holes
in Eq.\ref{hopping} is given by
\begin{eqnarray}
H_t &=& -{t_1 \over N} \sum_{\langle i,j \rangle}  
(f_i b^{\dagger}_{i,n\sigma} b_{j,n\sigma} f^{\dagger}_j + h.c.) \cr
&-& {t_2 \over N} \sum_{\langle \langle i,j \rangle \rangle}
(f_i b^{\dagger}_{i,n\sigma} b_{j,n\sigma} f^{\dagger}_j + h.c.)
\label{hoppingN}
\end{eqnarray}
Fourier transformation of these terms leads to
\begin{equation}
H_t=-{1 \over N}\sum_{\bf{k}\bf{k'}\bf{q}}
\G_{\bf{k}-\bf{k'}} f_{\bf{k}}b^\dag_{{\bf k'},n\s}b_{{\bf k'-q},n\s}f^\dag_{\bf{k-q}}
\label{Hhop}
\end{equation}
where
\begin{equation}
\G_{\bf k} = 2\square{t_1\round{\cos k_x+\cos k_y}+t_2\cos (k_x+k_y)}.
\label{Gamm a}
\end{equation}
These hopping terms give rise to the interaction between the hole and the
spin background.

In the LRO phase, the existence of the condensed bosons,
$|\langle b_{{\bf k},n\sigma} \rangle|^2 = N \delta^n_1 |x|^2
\delta_{{\bf k},{\bf Q}}$, leads to
\begin{eqnarray}
H_t &=& 2 \sum_{\bf{k}} \G_{\bf{k}-{\bf Q}} |x|^2
f^\dag_{\bf{k}}f_{\bf{k}} \cr
&+& {1 \over N} \sum_{\bf{k}\bf{k'}\bf{q}}
\sum_{m=2}^{N}\G_{\bf{k}-\bf{k'}}
b^{\dag}_{{\bf k'},m\s}b_{{\bf k'-q},m\s}f^{\dag}_{\bf{k}}f_{\bf{k-q}}
\label{Hhopspnmod}
\end{eqnarray}
The second term corresponds to the interaction between
the uncondensed spinons and the fermion and gives rise to
the fermion self-energy corrections.

\begin{figure}
\begin{center}
\includegraphics*[241,634][364,724]{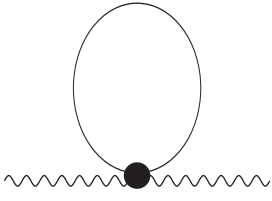}
\caption{\label{fig:bh} The self-energy correction of Eq.\ref{bh}.
The wavy line represents the fermionic propagator and the solid line
represents the bosonic propagator.}
\end{center}
\end{figure}

Let us first consider the perturbative self-energy corrections in
$t_1/N$ and $t_2/N$. The lowest-order correction is given by
(see Fig.3)
\begin{equation}
\S_1 ({\bf k}) = 2 \sum_{{\bf k'}}\G_{\bf k-k'}|v_{\bf k'}|^2,
\label{bh}
\end{equation}
where $(N-1)/N \approx 1$ is used.
The next order correction involves the diagram
in Fig.4 and can be computed as
\begin{equation}
\S_2({\bf k},\w) = {1 \over N} \sum_{{\bf k',q}}
f_{\mathbf{k} \mathbf{k'} \mathbf{q}}
G_f(\bf{k-q},\w-\e_{\bf{k'}}-\e_{\bf{q-k'}}),
\label{loop}
\end{equation}
where
\begin{equation}
f_{\mathbf{k} \mathbf{k'} \mathbf{q}}
= \left|\G_{\bf{k-k'}}u_{\bf{k'}}v_{\bf{k'-q}}+
\G_{\bf{k+k'-q}}u_{\bf{k'-q}}v_{\bf{k'}}\right|^2
\end{equation}
and $(N-1)/N^2 \approx 1/N$ is used.
Notice that, in the perturbative calculation, the bare fermionic
Green's-function, $G^0_f$, should be used for
$G_f$ in Eq.\ref{loop}.  

In order to take into account the multiple interactions, however,
we need to go beyond the perturbative evaluations of the self-energy
contributions. In our work, we will determine the renormalized fermion
Green's function, $G_f$, by summing up diagrams in the non-crossing
approximation. This amounts to using the renormalized Green's function
in Eq.\ref{loop} and solving the following Dyson equation for the fermion
Green's function.  
\begin{equation}
G^{-1}_f(\mathbf{k},\w)=
\w + 2 |x|^2 \G_{\bf{k}-{\bf Q}}-
\S_1 (\mathbf{k})-\S_2 (\mathbf{k},\w).
\label{scie}
\end{equation}
In the numerical solution of this equation, we set the coefficient of
the self-energy correction in Eq.\ref{loop} to be one for convenience
({\it i.e.} $1/N=1$). Notice that $2 |x|^2 \G_{\bf{k}-{\bf Q}}-\S_1 (\mathbf{k})$
corresponds to the dynamically generated 'band' dispersion for
the fermions at the lowest order.
In the case of the SRO phase, the same Dyson equation can be used
with $x=0$ in Eq.\ref{scie}.

\begin{figure}
\begin{center}
\scalebox{0.8}{\includegraphics*[188,577][413,709]{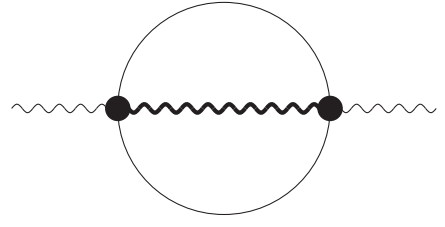}}
\caption{\label{fig:loop} The self-energy correction of Eq.\ref{loop}.
The thick wavy line is the exact fermionic propagator. Combined with
the contribution from Fig.\ref{fig:bh}, this gives rise to the
Dyson equation for the fermionic propagator in Eq.\ref{scie}.}
\end{center}
\end{figure}

\subsubsection{Evaluation of the Single-Hole Spectral Function}

In the LRO phase, the Green's function for the condensed bosons in the Fourier
space is given by
\begin{eqnarray}
G_{b,1\uparrow}({\bf k},\w) &=&
- N |x|^2 \de_{{\bf k},{\bf -Q}} \de (\w) \cr
G_{b,1\downarrow}({\bf k},\w) &=&
- N |x|^2 \de_{{\bf k},{\bf Q}} \de (\w).
\label{gblro}
\end{eqnarray}
On the other hand, the Green's function for the uncondensed bosons
are obtained as (for both $\sigma=\uparrow, \downarrow$)
\begin{eqnarray}
G_{b,m\s}({\bf k},\w+i\delta)=
\frac{|v_{{\bf k}}|^2}{(\w-\e_{\bf k}+i\delta)}
-\frac{u_{{\bf k}}^2}{(\w+\e_{{\bf k}}+i\delta)}.
\end{eqnarray}
Then the single-hole Green's function in Eq.\ref{hpf} can be written as
\begin{equation}
G_{h,\sigma} = G_{h,1\sigma} + \frac{N-1}{N}G_{h,2\sigma} \approx
G_{h,1\sigma} + G_{h,2\sigma}.
\end{equation}
The single-hole spectral function, $A_{h,\sigma}= -(1/\pi) {\rm Im} G_{h,\sigma}$
is now given by
\begin{equation}
A_{h,\sigma}({\bf k},\w) = A_{h,1\sigma}({\bf k},\w) + A_{h,2\sigma}({\bf k},\w).
\label{spectral_final}
\end{equation}
The first contribution from the condensed bosons is
\begin{eqnarray}
A_{h,1\uparrow}({\bf k},\w) &=&
-{1 \over \pi} |x|^2 {\rm Im} G_f({\bf k}+{\bf Q}, \omega+i\delta), \cr  
A_{h,1\downarrow}({\bf k},\w) &=&
-{1 \over \pi} |x|^2 {\rm Im} G_f({\bf k}-{\bf Q},\omega+i\delta).  
\label{cond}
\end{eqnarray}
The second term is due to the interaction between the fermion and
the uncondensed bosons and is given by
\begin{eqnarray}
&&\hskip -0.8cm A_{h,2\sigma} ({\bf k},\omega) \cr
&&\hskip -0.7cm =
-\sum_{\mathbf{q}} \int_{0}^{\w} dy A_f(\mathbf{q},y)
A_{b,2\sigma}(\mathbf{q-k},y-\omega) \hskip 0.2cm (\omega>0) \cr
&&\hskip -0.7cm =
\sum_{\mathbf{q}}\int_{\w}^{0}dy A_f(\mathbf{q},y)
A_{b,2\sigma}(\mathbf{q-k},y-\omega) \hskip 0.2cm (\omega<0),
\label{sfsro}
\end{eqnarray}
where $A_f=-(1/\pi){\rm Im}G_f$
and $A_{b,2\sigma}=-(1/\pi){\rm Im}G_{b,2\sigma}$.

On the other hand, in the spin-liquid state (or the SRO phase),
the first term in Eq.\ref{spectral_final} vanishes because there
is no boson condensate and the spectral function is entirely
given by the second term.

We will first numerically solve the Dyson equation for the fermion
Green's function in both the LRO and the spin-liquid phases.
Then the fermion spectral function can be convoluted with
the boson spectral functions in the two phases to
get the expression for the single-hole spectral function.
The evolution of the resulting single-hole spectral function
across the transition between the LRO and the spin-liquid phases
is discussed in the next section.  

\section{Numerical Results}

At first, the fermion Green's function is obtained by solving numerically
the self-consistent Dyson equation on a discrete mesh of ${\bf k}$ and $\w$ points.
In order to do so, we have solved the saddle point equation of the large-$N$ mean-field
theory for each set of $\kappa, J_1, J_2$, and use the mean field results as the
input to the Dyson equation of the fermion Green's function.  
The full hole Green's-function is then computed by using
Eqs.\ref{spectral_final},\ref{cond},\ref{sfsro}. The calculations are done with
a finite lattice of size $8^2$. We have confirmed that the overall structure
of the spectral function does not change when the size of the system is increased
to $16^2$.

\begin{figure}
\begin{center}
\includegraphics[bb=0 0 380 333,clip,scale=0.4]{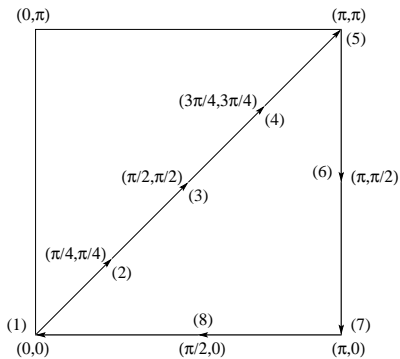}
\caption{\label{fig:bz} Eight momenta in the principal Brillouin zone at which the
spectral functions are evaluated.}
\end{center}
\end{figure}

\begin{figure}
\begin{center}
\includegraphics[bb=0 174 585 812,clip,scale=0.4]{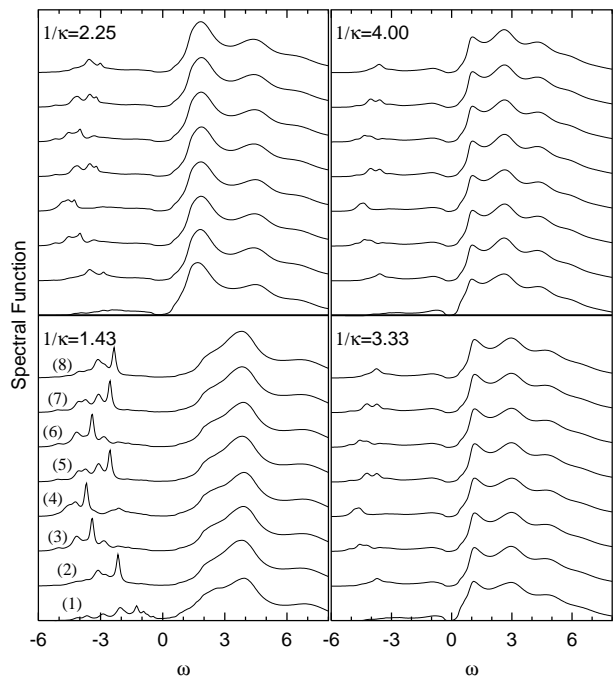}
\caption{\label{fig:esfs1} Evolution of the spectral functions along path 1 in Fig.
\ref{fig:triphase}. $J_1=J_2=1.0$ and $t_1/J_1=t_2/J_2=2.5$.}
\end{center}
\end{figure}

The evolution of the spectral function near the transition
between the incommensurate LRO and the spin-liquid phases, is examined at 8 different 
points in the phase diagram by following two different paths (see Fig.\ref{fig:triphase}). 
On both paths, the transition between two states is achieved by changing $\kappa$ 
for a given $J_2/(J_1+J_2)$. 
For each ground state, the spectral function is calculated at eight different momenta
in the principal Brillouin zone. These eight momenta are shown in Fig.\ref{fig:bz}.
When the spectral functions are plotted for these eight momenta, they will be arranged from
bottom up in the order indicated by the bracketed numbers as shown in Fig.\ref{fig:esfs1}.
All other similar plots will follow the same convention.

\begin{figure}
\begin{center}
\includegraphics[bb=0 172 585 812,clip,scale=0.4]{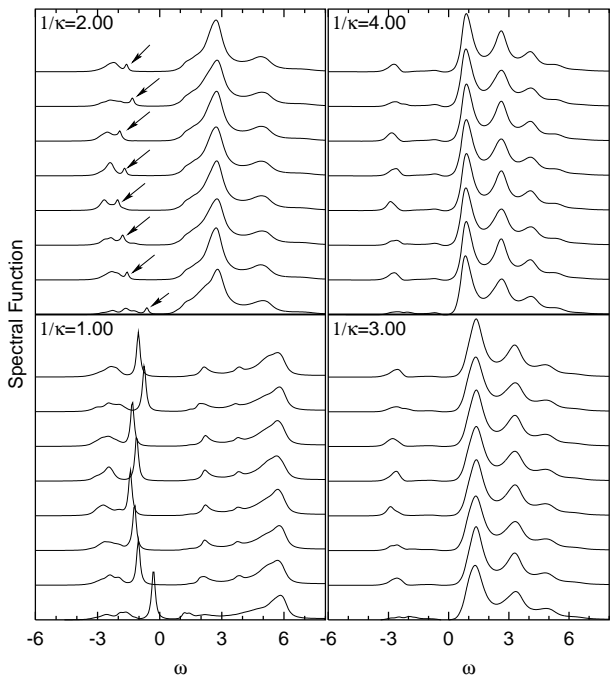}
\caption{\label{fig:esfs2} Evolution of the spectral functions along path 2 in Fig.
\ref{fig:triphase}. $J_1=0.75, J_2=1.0$ and $t_1/J_1=t_2/J_2=2.5$.}
\end{center}
\end{figure}

As explained earlier, quantum fluctuations are stronger for smaller values of $\kappa$.
Thus, the transition from the incommensurate LRO phase to the spin-liquid state can be
obtained by decreasing $\kappa$ for a given
value of $J_2/(J_1+J_2)$. We consider paths (1) and (2) corresponding to $J_1/J_2=1$
and $J_1/J_2=0.75$, respectively. The results for the four different
values of $\kappa$ are shown
in Fig.\ref{fig:esfs1} for path (1) and in Fig.\ref{fig:esfs2} for path (2).

The results of paths (1) and (2) are very similar. At large
$\kappa$, there exists a sharp low-energy
peak in the incommensurate LRO phase suggesting the existence of a coherent hole excitation
or a quasiparticle. Notice that the incoherent continuum at higher energies are due to
the existence of the 'unconfined' spinons. In the spin-liquid phase, the quasiparticle
peak completely disappears and only the incoherent broad continuum can be seen in the
spectral function.

Notice that the spectral weight of the quasiparticle peak in the
LRO phase decreases as the transition to the spin-liquid phase is approached and instead
the spectral weight of the incoherent continuum grows. This is due to the fact that the
spectral weight of the quasiparticle peak is proportional to the strength of the boson
condensate or the magnetic order parameter that decreases as the transition point is
approached. This can be observed in both Fig.\ref{fig:esfs1}
and Fig.\ref{fig:esfs2}, but is illustrated more clearly in the latter figure. In
Fig.\ref{fig:esfs2}, the strong quasiparticle peak seen when $\kappa =1$ diminishes
as we approach the transition. The remnant of the peak is indicated by the arrows in the
figure. Upon careful inspection, a slightly broad bump beside the peak consistently
persists in all of the plots indicating that
it is a contribution from the unconfined spinons. One can observe in both figures
that as the spectral weight of the quasiparticle peak shrinks, the remaining
spectral weight gradually moves to the 'unconfined' spinons at higher energies.

The incoherent continuum in the SRO phase arises because the injected hole
decays into its charge and spin degrees of freedom and they do not propagate
together in a coherent fashion. The origin of the broad continuum in the LRO phase
is basically the same; the 'unconfined' spinons make the motion of the hole
incoherent. The reason why the low-energy quasiparticle peak survives in the LRO phase
is that the density of states of the spinons becomes very small in the low energy
limit so that the spin degree of freedom of the injected hole at low energy cannot
efficiently decay into the spinon continuum.

We have also confirmed that varying the hopping amplitude, $t$,
does not change the qualitative behaviour of
the spectral function. The only change is that the bandwidth of
the quasiparticle is increased as the hopping amplitude becomes bigger.

\section{Summary and Conclusion}

We investigated the single-hole spectral function in the
anisotropic-triangular-lattice Heisenberg antiferromagnet.
In particular, we studied the evolution of
the single-hole spectral function across the transition between
the incommensurate LRO phase and the
spin-liquid state. This phase transition can be described by the
large-$N$ Sp($N$) mean-field theory\cite{read90,sachdev91,read91,sachdev91-2,sachdev92},
where the Sp($N$) analog of the spin magnitude, $\kappa$, can be adjusted
to control the strength of quantum fluctuations.
The motion of the single hole injected into the insulator is studied via the
$t$-$J$ type model in the slave fermion representation, where the electron is
represented as the composite of a charge-carrying fermion and a spin-carrying
boson. Taking into account the multiple interactions between the injected hole
and the excitations of the underlying insulator in the non-crossing approximation,
we evaluated the single-hole spectral function in both of the LRO and the spin-liquid
phases.
    
It is found that, in the LRO phase, the quasiparticle peak exists at low energy
and the broad incoherent background also arises at intermediate to high energies.
This incoherent continuum arises due to the existence of 'unconfined' spinons
in the large-$N$ mean-field theory. In principle, these spinons should be
confined below a 'confinement' energy scale due to a fluctuation effect beyond
the mean-field theory\cite{sachdev03}. {\it Near} the transition, however, the confinement
energy scale becomes very small (in fact it is zero at the transition) so that
the spinons at intermediate to high energies are 'almost' unconfined.
In this respect, our calculation is particularly suited to describe the
spectral function near the transition. As the transition to the spin-liquid
state is approached, the spectral weight of the quasiparticle peak goes to
zero and the incoherent continuum grows. Finally, in the spin-liquid phase,
only the incoherent continuum survives in the spectral function.
This is due to the fact that the injected hole can decay into the
charge-carrying fermionic excitation and the spin-carrying spinon
and the motions of these excitations are not coherent together.

In our studies, we neglected various vertex corrections, which may have
lead to an underestimation of the coherence, especially in the spin-liquid
phase. However, the recent exact diagonalization study of the spin-liquid
state on the Kagom\'e lattice showed spectral functions that were indeed
completely incoherent\cite{lauchli03},
leading to the conclusion that the vertex correction would not change the
qualitative behaviours obtained in our study.

Our work is relevant to possible ARPES experiments on Mott insulators
on the anisotropic triangular lattice since the ARPES on the insulator
can measure the single-hole spectral function\cite{wells95}.
In particular, notice that the neutron
scattering experiment on Cs$_2$CuCl$_4$ reveals the existence of the spinon
continuum at intermediate to high energies in the LRO phase near the transition
as well as in the spin-liquid phase\cite{coldea03,coldea01}.
The parameter $1/\kappa$ in the large-$N$ theory plays the same role as
the in-plane magnetic field in the experiment on Cs$_2$CuCl$_4$, where the
transition between the two phases is achieved by changing the in-plane field or
temperature. Thus we expect that, in light of our work, the ARPES experiments
on such systems will provide independent confirmation of the spinon continuum
discovered in the neutron scattering experiments on Cs$_2$CuCl$_4$ or in related
systems.

{\bf Acknowledgment}: This work was supported by the NSERC of Canada,
Canadian Institute for Advanced Research, and Canada Research Chair Program.

\end{document}